\documentclass[aps,prl,showpacs,preprint]{revtex4}

\usepackage{graphicx,amsmath}
\usepackage{wasysym,pifont}
\usepackage{color}

\begin{document}

\newcommand{\be}{\begin{equation}}
\newcommand{\ee}{\end{equation}}
\newcommand{\bea}{\begin{eqnarray}}
\newcommand{\eea}{\end{eqnarray}}
\newcommand{\<}{\langle}
\renewcommand{\>}{\rangle}
\newcommand{\qea}{q_{\scriptscriptstyle{\rm EA}}}
\newcommand{\teff}{t_{\rm eff}}

\title{
Edwards-Anderson spin glasses undergo simple cumulative aging}

\author{Andrea Maiorano}
\author{Enzo Marinari}
\author{Federico Ricci-Tersenghi}
\affiliation{Dipartimento di Fisica, SMC-INFM and INFN\\
Universit\`{a} di Roma ``La Sapienza'', P. Aldo Moro 2, 00185 Roma,
Italy}

\date{\today}

\begin{abstract}
We study and discuss rejuvenation and memory (numerical) experiments
in Ising and Heisenberg three and four dimensional spin glasses. We
introduce a quantitative procedure to analyze the results of
temperature cycling experiments. We also run, compare and discuss
``twin'' couples of experiments.  We find that in our systems aging is
always cumulative in nature, and rejuvenation and memory effects are
also cumulative: they are very different from the ones observed in
experiments on spin glass materials.
\end{abstract}

\pacs{75.50.Lk,64.70.Pf}

\maketitle

Rejuvenation and memory effects (that in the following we will denote
by RME) are maybe the most striking features of spin glass materials,
and their experimental evidence is very clear (see for example the
discussion in \cite{MIYVIN} and references therein).  The simplest
experiment showing both rejuvenation and memory in a spin glass is the
cycle in temperature, which is based on the following three steps:
\begin{enumerate}
\item 
one lets a spin glass sample relaxing for a time $t_1$ at temperature
$T_1$ in the low $T$ phase;
\item
then one brings it to a temperature $T_2<T_1$, where it relaxes for a
time $t_2$;
\item
finally one heats it back to $T_1$ where relaxation continues.
\end{enumerate}
The relevant experimental observations are mainly the following two:
\begin{itemize}
\item 
independently from the amount of time spent at $T_1$, when the sample
is cooled to $T_2$ the relaxation process restarts completely {\em
(rejuvenation)};
\item
when the sample is heated back to $T_1$ it seems to remember what
happened during time $t_1$ and relaxation continues as if the second
step was absent {\em (memory)}.
\end{itemize}
We have given here a very simplified description, but it is sufficient
for our goal, that is to compare the situation to the results of
numerical simulations.  Actual experiments show a number of different
and subtle effects; we address the interested reader to the
experimental results of \cite{MIYVIN} (and references therein).

RME effects are poorly understood from the theoretical point of view:
for example it is still unclear which are the length scales that are
relevant in such processes.  Unfortunately length scales can not be
measured directly in experiments, and numerical simulations could be
of great help in this context.  At the best of our understanding it is
not clear, today, if real RME (of the same nature of the ones observed
in experiments) appear in numerical simulation of finite dimensional
Edwards-Anderson (EA) models (with either Ising or Heisenberg spins).
In this note we clarify this point.  We use a phenomenological
approach: rather than trying to interpret numerical data within a
specific theory in order to validate it we focus on the comparison of
numerical and experimental data.  Our aim is to check whether RME, as
observed in physical experiments, are also present in the EA model. In
order to reach conclusions as general as possible, we consider EA
models with different spin types (Ising and Heisenberg) and in $D=3$
and $D=4$.

Let us start with a brief review of RME as observed (or not) in
numerical simulations of the EA model.  A few years ago the work of
\cite{KOTAYO,PIRIRI} discussed numerical studies of such effects in
the $3D$ Ising EA model with Gaussian couplings.  Unfortunately the
lack of a quantitative method for estimating RME brought the authors
of the these two studies to give different interpretations of the
outcome of a ``cooling and stop'' experiment.  While reference
\cite{KOTAYO} states that "the model exhibits the rejuvenation-like
and memory effects within a time-window of the present simulation",
reference \cite{PIRIRI} says that "the model does not show, on the
time scales we have access to, the strong RME real spin glasses show"
(the time scales of the two numerical experiments are of the same
order of magnitude, and the spatial volumes of the two systems are
comparable).

A couple of years later, two further numerical works on this issue
\cite{BERBOU,HUKTAK} reach again opposite conclusions.  Berthier and
Bouchaud~\cite{BERBOU} interpret their data for the $4D$ Ising EA
model as showing strong RME.  They also suggest that in $3D$ such
effects are difficult to observe because the spatial correlation
function does not change enough when varying the temperature.  On the
contrary Takayama and Hukushima~\cite{HUKTAK} find the signature of a
{\em cumulative aging} scenario for small $\Delta T\equiv T_1-T_2$.
The cumulative aging scenario assumes that, as long as the system is
in a spin glass phase, temperature changes do not induce a restart of
aging, so that effects of relaxations at different temperatures
cumulate.

In a recent paper Jimenez \textit{et al.}~\cite{JIMAPE} find again RME
in both $3D$ and $4D$ Ising EA models.

Given such a confusing situation and such a number of different
numerical results, we have decided to make very precise measurements
with temperature cycle experiments in the EA model in order to try to
answer the following three questions:
\begin{enumerate}
\item
are true RME present in the EA model or is aging cumulative in nature?
\item 
if we observe cumulative aging, can we try to understand if true RME
can be recovered in the limit of very large (relaxing and probing)
time scales, i.e.\ in the limit relevant for experiments?
\item 
how much these effects depend on space dimension and spin type?
\end{enumerate}
We consider the EA model with Gaussian couplings and both Ising spins
(in $3D$, I3D, and $4D$, I4D) and Heisenberg spins (in $3D$, H3D).
Typical sizes used are $L=40$ for I3D, $L=20$ for I4D and $L=60$ for
H3D.  We have checked that our lattices are large enough to avoid any
detectable finite size effect; in particular, the choice of a large
size for H3D samples is due to the very large length scales involved
in the dynamics of Heisenberg spin glasses \cite{BY}.  We have
computed the disorder averages by using $16$ samples for H3D, $176$
samples for I4D and $256$ samples for I3D. The dynamics of Ising
spins models has been based on the popular single spin-flip Metropolis
algorithm.  For Heisenberg spins we use again Metropolis updates but
when changing temperature we fix the acceptance ratio, so the
amplitude of the trial updates depends on the temperature: in this way
we reproduce at best the physical dynamics.  Most of the numerical
simulations of I3D were performed on the APEmille parallel computer
\cite{APEMIL}, while I4D and H3D were simulated on a PC cluster.

The choice for $T$-cycle experiments, which are in principle more
complicated than $T$-shift experiments, is dictated by two main
reasons. First, $T$-cycle experiments allow for the study of both
rejuvenation and memory at the same time. Second, the first part of
any relaxation process may be affected by large finite time
corrections.  Thus it is important that, when extracting the effective
age of the system (see below), one compares relaxation processes where
the initial steps are performed at the same temperature. In other
words the effective age must be measured deep into the aging regime
and not in the initial part of the relaxation process.

In order to simplify the analysis, especially when taking the large
time limit, we introduce in our experimental procedure only one single
relevant time scale $t_p$, that corresponds to the period of the
measuring field used in real experiments.  $t_p$ is the
number of MC steps on which we average data: in other words we divide
the total number of MC steps in groups of $t_p$ steps over which we
compute expectation values. We use $t_p$ also as the time distance for
computing time dependent correlations over spin configurations.  We
perform different runs for each fixed value of $t_p=200, 500, 1000,
2000, 5000$ (I3D), $t_p=100,1000$ (I4D, H3D).  All the other time
scales will be proportional to $t_p$.  In particular, if $t_i$ is the
time spent in the phase $i$ of the experiment, we fix $t_1 = t_2 = t_3
= 20 t_p$.

Our first aim is to define properly an effective time $\teff$, such
that after a $T$-cycle (i.e.\ $t_1$ steps at $T_1$, $t_2$ at $T_2$ and
$t_3$ at temperature $T_1$ again) the system is in the same state as
if it was let relaxing isothermally at $T_1$ during the time
$t_1+\teff+t_3$. Because of possible transient effects (just after
restoring temperature $T_1$) one should avoid to use small values of
$t_3$.

Checking that two systems are statistically equivalent is not easy.
We have done that by comparing a number of observable quantities and
checking whether their values coincide in our statistical accuracy.
We have considered both one time and two time quantities.  As one time
quantities we look at the Edwards-Anderson overlap order parameter
$\qea$ and the spatial correlation function $G(x,t)$:
\begin{equation}
\label{eq:qea}
\qea(t) \equiv \frac{1}{N}\sum_{i=1}^N\overline{m_i(t) \cdot
m_i(t)}\\
\end{equation}
\begin{equation}
\label{eq:gxt}
G(x,t) \equiv
\frac{1}{zN}\sum_{\|i-j\|=x}\overline{\left(S_i^a\left(t\right) \cdot
S_j^a\left(t\right)\right) \left(S_i^b\left(t\right) \cdot
S_j^b\left(t\right)\right)}
\end{equation}
where $\overline{\ \cdot\ }$ denotes the average over the quenched
disordered couplings and thermal histories, $z$ is the coordination
number of the simple cubic $D$-dimensional lattice, $N=L^D$, and $a$,
$b$ are real replica indexes.  The basic fields of the theory take
values $S_i=\pm1$ variables for the Ising spin glasses (I3D, I4D),
while are vectors on a sphere of unitary radius in the case of the
Heisenberg spin glass (H3D). We define the corresponding
time-integrated magnetization (over the time $t_p$) as
\begin{equation}
\label{eq:mi}
m_i(t) \equiv \frac{1}{t_p}\sum_{\tau=t-t_p+1}^{ t}S_i(\tau)
\end{equation}
We have also measured and used some two time quantities: the in-phase
and out-of-phase susceptibilities. Provided that we are in the
quasi-equilibrium regime (so that FDT holds) they can be estimated via
the spin autocorrelation function
\begin{equation}
\label{eq:cnormal}
C(t,t') \equiv \frac{1}{N} \sum_{i=1}^N \overline{S_i(t) \cdot S_i(t')}
\end{equation}
In order to improve the signal-to-noise ratio we have integrated the
above autocorrelation function over short times, $\tau <t_p$, and
defined
\begin{equation}
\label{eq:ctilde}
\tilde{C}(t,t') \equiv \frac{1}{N} \sum_{i=1}^N \overline{m_i(t) \cdot
m_i(t')}
\end{equation}

The in-phase and out-of-phase
susceptibilities are then expressed as a function of this
time-integrated correlation function
\begin{eqnarray}
\label{eq:chip}
\tilde{\chi}^\prime(t,t+t_p) & \equiv & \frac{\tilde{C}(t,t) -
\tilde{C}(t,t+t_p)}{T}\ \ \ ,\\
\label{eq:chis}
\tilde{\chi}^{\prime \prime}(t,t+t_p) & \equiv &
\frac{1}{T} \left[\tilde{C}(t,t) + \tilde{C}(t-t_p,t+t_p)\right. \nonumber \\ &
 - &
\left. \tilde{C}(t-t_p,t) - \tilde{C}(t,t+t_p)\right]\ \ \ .
\end{eqnarray}
Since $\tilde{\chi}^{\prime \prime}$ showed too small excursions upon
$T$ changes, we preferred to use the out-of-phase susceptibility
defined via the spin-spin autocorrelation (\ref{eq:cnormal})
\begin{eqnarray}
\label{eq:chisnormal}
\chi^{\prime \prime}(t,t+t_p) & \equiv &
\frac{1}{T} \big[{1 + C(t-t_p,t+t_p)} \nonumber \\ 
& - & {C(t-t_p,t) - C(t,t+t_p)}\big]
\end{eqnarray}
Notice that relations (\ref{eq:chis}) and (\ref{eq:chisnormal}) hold
apart from an overall multiplicative factor \cite{KOTAYO}.

For each period $t_p$ we measure one-time quantities only at the end
of period, and we integrate measurements of two-time quantities over
the period, in close analogy with real experiments.  This observation
can be relevant since the only data which have been interpreted as a
rejuvenation effect in \cite{BERBOU} have been measured before the end
of the first period after the $T$-shift: this is not done in real
experiments.  A possibility that we consider plausible is indeed that
the RME showed in \cite{BERBOU}
are not related to the experimental RME effects, but to the fact that
right after the temperature change the system is (for a very short
time) strongly out of equilibrium. In such a situation
the response measured in \cite{BERBOU} with the expression
$\chi(t)\equiv\frac{1-C(t,t+t_p)}{T}$ may overestimate the true
susceptibility, giving rise to a signal which looks like a stronger
rejuvenation.  A deeper analysis of this effect has been done in
\cite{JIMAPE}.

\begin{figure}
\includegraphics[width=\columnwidth]{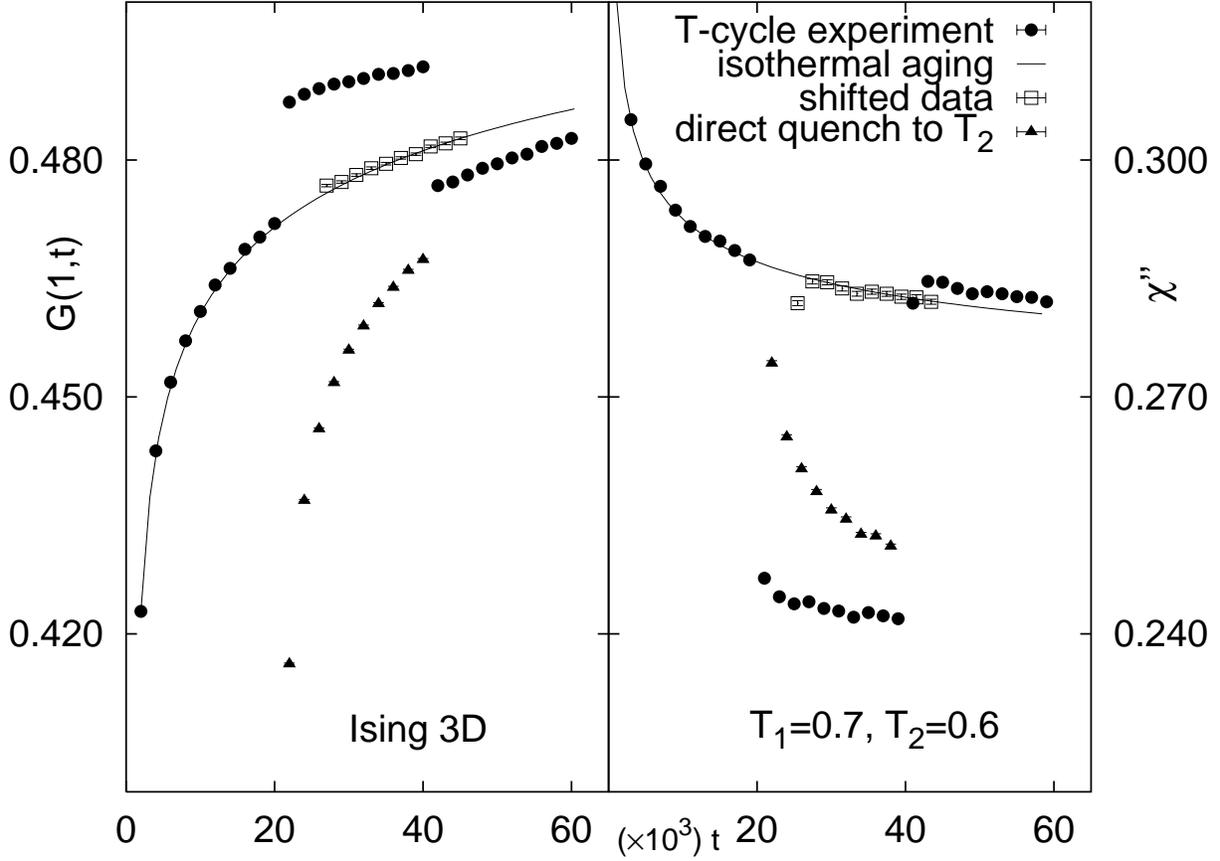}
\caption{Spatial correlation function at distance one (that coincides
with the link overlap) and out-of-phase susceptibility (as defined in
the text) in a temperature cycle of I3D. The time evolution of
$G(1,t)$ and $\chi^{\prime \prime}$ are compared with the evolution
coming from a direct quench at $T_2$. The continuous line is a fit on
data from a long isothermal run at $T_1$.}
\label{fig:i3d_c4_chi2}
\end{figure}

\begin{figure}
\includegraphics[width=\columnwidth]{i4d_c4_chi.eps}
\caption{As in figure \ref{fig:i3d_c4_chi2} for I4D, but
$\tilde{\chi}^{\prime}$ instead of $\chi^{\prime\prime}$.}
\label{fig:i4d_c4_chi}
\end{figure}

Figures \ref{fig:i3d_c4_chi2} and \ref{fig:i4d_c4_chi} show how our
method for estimating $\teff$ works.  In Fig.~\ref{fig:i3d_c4_chi2} we
show $G(1)$ and $\chi^{\prime\prime}$ for I3D in a $T_1=0.7$,
$T_2=0.6$ cycle (remember that here $T_c\simeq 0.95$), while in
Fig.~\ref{fig:i4d_c4_chi} we show $G(1)$ and $\chi'$ for I4D in a
$T_1=1.3$, $T_2=0.9$ cycle (here $T_c\simeq 1.8$).  For clarity only
half of the data points are presented in the figures. 
In each plot we show raw data measured during the
$T$-cycle ($\bullet$). Isothermal aging data, from very long
($300t_p$) simulations at fixed $T=T_1$, are fitted on a simple smooth
function $f(t)$ (that fits perfectly the data and is only used as a
book keeping device for the matching procedure) that we represent with
a solid line.  $\teff$ is calculated by shifting horizontally $f(t)$
to fit the data from the third stage of the $T$-cycle, and adding the
needed time shift $t_s$ to the time $t_2$. $t_s$ enters the procedure
as a fitting parameter, allowing a fully automatized estimation of
$\teff=20t_p+t_s$, so we do not introduce any systematic error due to
human perception of collapsing goodness.

In these plots we do not see a real and complete rejuvenation
as in experiments, where the susceptibility decays in the
second stage as if the first stage was absent (at least for a large
$\Delta T$ as the one we are using here).  This is clear especially
if we compare the second part of the $T$-cycle with a direct quench at
$T_2$ ({\small\ding{'163}}). The authors of \cite{BERBOU} suggested
that in real experiments even the fastest quench always corresponds to
a cooling, so that the starting configuration of any relaxation
process is never completely random.  In order to check how much this
fact could affect our hypothesis that the relaxation at $T_2$ strongly
depends on the time spent at $T_1$, we have computed a new direct
quench curve, starting this time not from a completely random
configuration ($T=\infty$), but from a configuration thermalized at
temperature $T = 2 T_c$.  Again we find substantial differences
in the observables decays between these {\em softer}
direct quenches and the relaxation in the second stages of the
$T$-cycles: 
the two decays are not the same, and the discrepancy is not too
different than the one from a direct quench (see
Figures \ref{fig:i3d_c4_chi2} and \ref{fig:i4d_c4_chi})
This shows that even starting from a slightly correlated
spin configuration, that is what could be happening in real
experiments, we do not recover the behavior of the temperature cycle.

\begin{figure}
\includegraphics[width=\columnwidth]{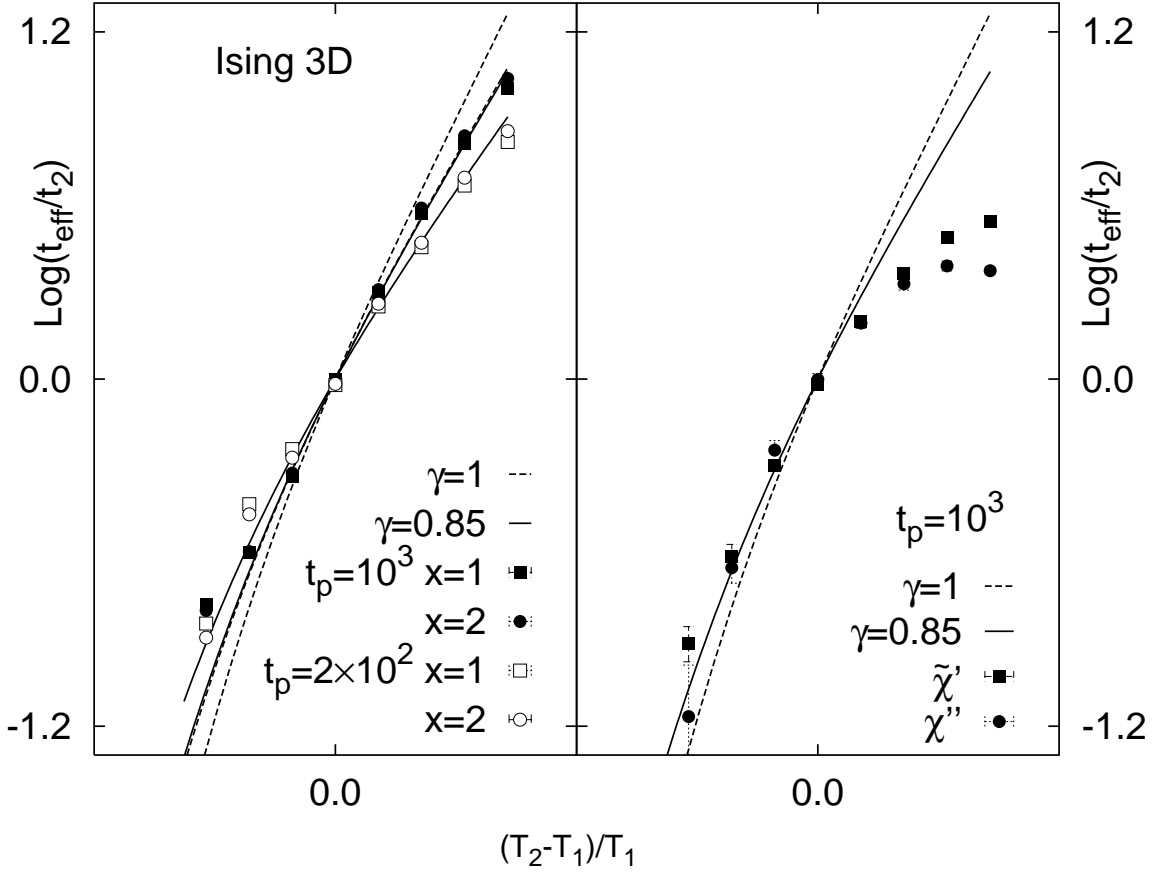}
\caption{Comparison between the ratio $\frac{\teff}{t_2}$ and the
cumulative hypothesis, for I3D. Measurements of $\frac{\teff}{t_2}$
are from cycle measurements of $G(x,t)$ with $x=1,2$ (left frame) and
of the susceptibilities (right frame). In the left frame filled
symbols are from measurements in cycles with $t_p=1000$, while empty
symbols come from cycles with $t_p=200$. These plots are
representative of the behavior of one-time and two-times
quantities. Full and dashed lines are the theoretical prediction in
the cumulative hypothesis with $\gamma=0.85$ and $\gamma=1$
respectively (on the right part of the plot higher lines have a larger
$t_p$ value).}
\label{fig:tweff_i3d_1t_2t}
\end{figure}

Having estimated $\teff$ for different values of $T_2$ we summarize
our results in figure \ref{fig:tweff_i3d_1t_2t}.  As already noticed
for example in \cite{PIRIRI} (see also \cite{BERBOU,HUKTAK}), positive
$\Delta T$ cycles do not reinitialize aging as in real experiments. On
the contrary the time spent at $T_2>T_1$ strongly increases the
relaxation rate: for $\Delta T>0$, $\teff$ is larger than $t_2$. Full
and dashed lines in figure \ref{fig:tweff_i3d_1t_2t} correspond to
predictions obtained in a fully cumulative aging scenario (see
below). $\teff$ values are very far from experimental observations,
which predict $\teff=0$ for $\Delta T<0$ and $\teff=-t_1$ for $\Delta
T>0$.  Moreover the experimental behavior does not seem to be
approached when we let the simulation time scales grow.  We show in
the left frame of figure \ref{fig:tweff_i3d_1t_2t} the data for
$\teff$ obtained with $t_p=1000$ and $t_p=200$. Both of them are well
described by the cumulative hypothesis: this suggests that the
cumulative aging scenario remains valid for very long ages of the
system.  In other words this ``cumulative'' behavior does not change
when changing the total duration of the experiment by modifying the
value of $t_p$.  In the right frame of the same figure we show the
same measurements obtained from two-times observables.  They do not
appear to be in good agreement with the cumulative hypothesis for
$T_2>T_1$ where our data points flatten: relaxation of these two time
observables is much flatter in the third stage of the cycle
(especially for large positive $T$ shifts) and the fitting procedure
to estimate $\teff$ is affected by a very large incertitude.

Let us discuss how we have obtained the analytical predictions shown
in figure \ref{fig:tweff_i3d_1t_2t}.  We assume that the
off-equilibrium correlation length grows as $\xi_T(t)$ (a $T$
dependent functional dependence over time). In this case the
cumulative aging prediction for $\teff$ is that
\begin{equation}
\label{eq:xi}
\xi_{T_1}(t_1+\teff) =
\xi_{T_2}\Big(\xi^{-1}_{T_2}\big(\xi_{T_1}(t_1)\big)+t_2\Big)\ .
\end{equation}
The correlation length in Ising EA models is believed to grow as
\cite{MPRR} $\xi_T(t) \propto t^{A}$, with $A=aT$ (with $a\sim 0.17$
in $3D$).  Using this functional dependence we obtain the dashed line
in figure \ref{fig:tweff_i3d_1t_2t}.  We also explore the possibility
of a more general dependence \cite{Futuro} by assuming that
$A=aT^\gamma$ (the usual dependence assumes $\gamma=1$).  The best fit
to new high-precision data \cite{Futuro} gives $\gamma=0.85\pm 0.04$.
With $\gamma=0.85$ the prediction for $\teff$ becomes the one plotted
with a full line in Fig.~\ref{fig:tweff_i3d_1t_2t}, which is a much
better interpolation of the numerical data.

\begin{figure}
\includegraphics[width=\columnwidth]{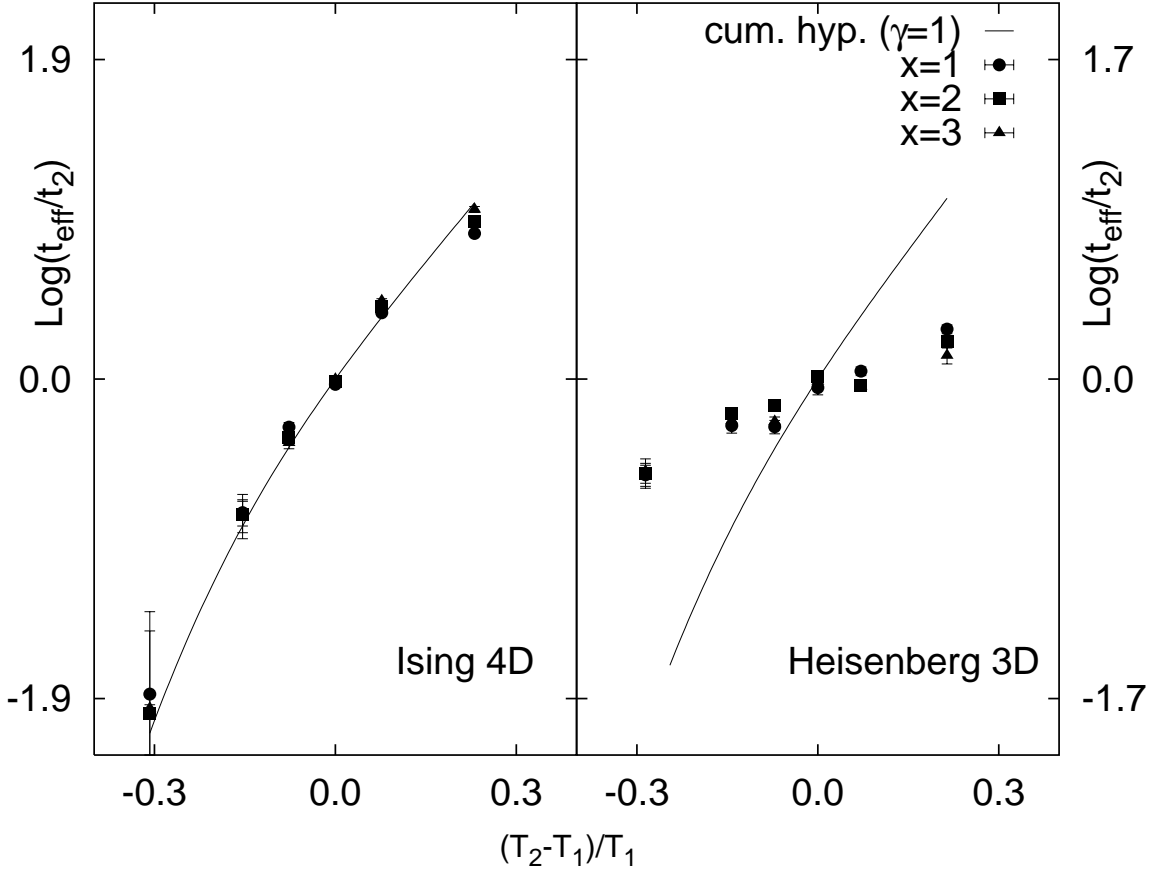}
\caption{As in figure \ref{fig:tweff_i3d_1t_2t}, but for I4D and H3D.
Values are extracted from measurements of $G(x,t)$, $x=1,2,3$, $t_p=1000$.}
\label{fig:tweff_i4d_h3d}
\end{figure}

In figure \ref{fig:tweff_i4d_h3d} we present the analogous data for
I4D and H3D, and we compare them with a cumulative hypothesis based on
$\gamma=1$ (this is only to allow to compare to a scenario where $\xi$
grows as a power of the time: at least for the Heisenberg case we have
no precise hints about a given rate of growth, so that the fact that
the solid curve does not fall on the numerical data cannot be seen as
a ``discrepancy''). Data for H3D show a very weak dependence of
$\teff$ (and of $\xi(t)$) on $T$, which deserves (and is undergoing)
deeper investigations \cite{Futuro}.

The authors of \cite{HUKTAK} find cumulative aging only for small
$\Delta T$, while for $\Delta T \ge 0.3$ their data are incompatible
with the cumulative aging scenario. This incompatibility shows up as
an asymmetry in the laws for transforming times from $T_1$ to $T_2$
and that from $T_2$ to $T_1$.  In the cumulative aging scenario these
two functions should be one the inverse of the other, while in
\cite{HUKTAK} they are shown not to be so.  This discrepancy could be
due to the fact that $T$-shift experiments of \cite{HUKTAK} also take
into account the very first part of the relaxation after the initial
quench, which is typically plagued by finite time effects.  We believe
that in order to avoid this kind of problems any measurement should be
taken late enough after the initial quench, in such a way that the
system has already entered the asymptotic aging regime (and in any
case all the region of very large $\Delta T$ is bound to be affected
by non-universal effects, very resilient to a clean theoretical
analysis).

In order to investigate this potential problem we have repeated the
``twin-experiments'' of \cite{JONOYO,HUKTAK}.  They are based on four
stages: the first stage ($t_1$ steps at $T_1$) is the same in both
twin experiments, and it is only used to bring the system in the
asymptotic aging regime (in this stage there are no measurements). In
the following two stages the two experiments are complementary: one
consists of $t_2$ steps at $T_2$ and then $t_3$ steps at $T_3=T_1$,
while the other goes first with $t'_3$ steps at $T_3=T_1$ and then
$t'_2$ steps at $T_2$. In the fourth and last stage both experiments
are run at the same temperature $T_4=T_1$.

\begin{figure}
\includegraphics[width=\columnwidth]{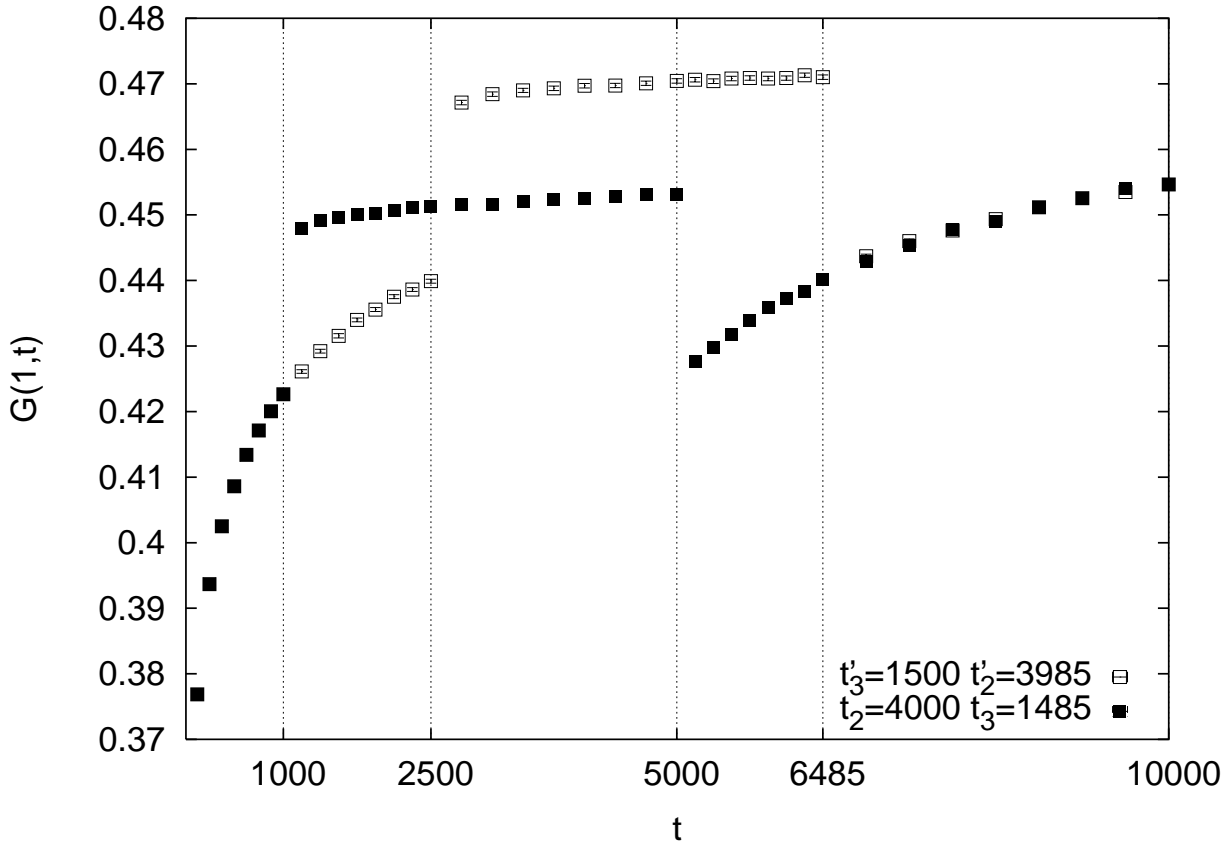}
\caption{Twin T-cycle experiment ($T_1=0.7$, $T_2=0.4$) of I3D, based
on measurements of $G(1,t)$. Stage durations were predicted using a
power law cumulative hypothesis and imposing the equivalence of the
coherence length at the end of the third stage.}
\label{fig:twin_i3d_0.7_0.6}
\end{figure}

Assuming the validity of the cumulative aging hypothesis, it is not
difficult to choose times $t_1$, $t_2$, $t'_2$, $t_3$ and $t'_3$ at
fixed temperatures $T_1$ and $T_2$ such that the correlation length
takes the same value at the end of the complementary stages (the
second and the third ones).  If the cumulative aging hypothesis is
correct, one should observe that in the fourth stage measurements from
the twin experiments coincide.  We show in figure
\ref{fig:twin_i3d_0.7_0.6} the results of measurements of $G(1)$ from
twin experiments of I3D with $T_1=0.7$, $T_2=0.4$: stage durations are
marked by dotted vertical lines.  In the fourth stage measurements of
$G(x)$ turn out to coincide (in our statistical accuracy) even for
large values of $x$: the structures built by the two system undergoing
different histories are, as far as we can check, equivalent.

We believe that we have been able to give {\em quantitative} evidence
that aging in finite dimensional spin glasses is {\em cumulative} in
nature. It is clear that, as always in numerical simulations, our
statements are valid in the limit of, among others, the time scales we
are able to investigate (that are far shorter than the experimental
ones).  Still, our search for some potential asymptotic restoration of
true, experimental-like RME has failed.  Our findings concern both
Ising and Heisenberg systems, both in $3D$ and in $4D$: in our time
windows the behavior is not substantially affected under a sizable
change of time window, even if we should not forget that we are still
very far from the experimental time scales.

It is clear that further studies are required. There is a clear
mismatch with experimental data, where a non-trivial aging is
observed.

\acknowledgments

This work was partially supported by the European Community's Human
Potential Programme under contracts HPRN-CT-2002-00307, Dyglagemem,
and HPRN-CT-2002-00319, Stipco.

\end{document}